\documentclass[reprint,aip,numerical,amsmath,fleqn,groupedaddress]{revtex4-1}
\usepackage[mathlines]{lineno}
\modulolinenumbers[1]
\usepackage{graphicx}
\usepackage[hidelinks]{hyperref}

\begin{document}

\preprint{arXiv:1211.0697}

\title[Plastic materials for carbon-ion radiotherapy]{Evaluation of plastic materials for range shifting, range compensation, and solid-phantom dosimetry in carbon-ion radiotherapy}

\thanks{This study has been published in Medical Physics 40(4) 041724, April 2013 (\url{http://dx.doi.org/10.1118/1.4795338}).}

\author{Nobuyuki Kanematsu}\email[]{nkanemat@nirs.go.jp}
\author{Yusuke Koba}
\author{Risa Ogata}
\affiliation{Research Center for Charged Particle Therapy, National Institute of Radiological Sciences, 4-9-1 Anagawa, Inage-ku, Chiba 263-8555, Japan}

\date{November 2012---April 2013, Reprint: \today}

\begin{abstract}
\begin{description}
\item[Purpose]
Beam range control is the essence of radiotherapy with heavy charged particles.
In conventional broad-beam delivery, fine range adjustment is achieved by insertion of range shifting and compensating materials.
In dosimetry, solid phantoms are often used for convenience.
These materials should ideally be equivalent to water.
In this study, we evaluated dosimetric water equivalence of four common plastics, HDPE, PMMA, PET, and POM.
\item[Methods] 
Using the Bethe formula for energy loss, the Gottschalk formula for multiple scattering, and the Sihver formula for nuclear interactions, we calculated the effective densities of the plastics for these interactions. 
We experimentally measured variation of the Bragg peak of carbon-ion beams by insertion of HDPE, PMMA, and POM, which were compared with analytical model calculations.
\item[Results]
The theoretical calculation resulted in slightly reduced multiple scattering and severely increased nuclear interactions for HDPE, compared to water and the other plastics.
The increase in attenuation of carbon ions for 20-cm range shift was experimentally measured to be 8.9\% for HDPE, 2.5\% for PMMA, and 0.0\% for POM while PET was theoretically estimated to be in between PMMA and POM.
The agreement between the measurements and the calculations was about 1\% or better. 
\item[Conclusions]
For carbon-ion beams, POM was dosimetrically indistinguishable from water and the best of the plastics examined in this study.
The poorest was HDPE, which would reduce the Bragg peak by 0.45\% per 1-cm range shift, although with marginal superiority for reduced multiple scattering.
Between the two clear plastics, PET would be superior to PMMA in dosimetric water equivalence.
\end{description}
\end{abstract}


\maketitle

\section{Introduction}

The essence of radiotherapy with heavy charged particles is its intrinsic capability of three-dimensional dose formation with a Bragg peak at the beam range, which can be precisely controlled by adapting the beam energy incident on a patient.
In conventional broad-beam delivery,\cite{Koehler 1977, Kanai 1999} a range shifter degrades excessive beam energy and a range compensator fills target-depth deficit varying in the field to best conform the spread-out Bragg peak to a planning target volume.\cite{Wagner 1982}

A range shifter is typically a composite of plates of precisely controlled thickness and a range compensator is a physical filter custom-made for an individual field.
Their material should be uniform, stable, machinable, inexpensive, and ideally equivalent to water, which is the reference material for dosimetry.
These requirements are also valid for solid phantom materials, which are used when water-phantom dosimetry is not possible or convenient.
To estimate the effective thickness of a material by the resultant range shift in water, an effective density is assigned for the material.
However, this approximation may cause dosimetric errors due to alteration of the radiation because the effective density differs for multiple scattering and nuclear interactions.\cite{Kanematsu 2012}
Dosimetric water equivalence, which is evaluated by uniformity of effective densities for relevant interactions, is essential to energy degrading of charged-particle beams for range shifting, range compensation, and dosimetry.

Some common materials used for this purpose are high-density polyethylene (HDPE),\cite{Kanematsu 2007} polymethyl methacrylate (PMMA),\cite{Wagner 1982} and synthetic resin of acrylonitrile, butadiene, and styrene (ABS).\cite{Takada 2008}
In general, HDPE is inexpensive and close to water in density, PMMA is good at dimensional stability and contains oxygen which is dominant in water, and ABS is available in various forms in industry as chemical wood. 
For proton and ion-beam dosimetry, PMMA has been commonly used as a water-equivalent phantom material.\cite{Brusasco 2000, Yajima 2009}

In the past, water equivalence of general plastics (PMMA, HDPE, PS, and PTFE), tissue-substitute plastics (A150 and commercial products), and other solid materials (graphite, bone, aluminum, and copper) have been studied for energy degrading and calorimetry of proton and ion beams by experiment and Monte Carlo simulation.\cite{Palmans 2002, Schneider 2002, Al-Sulaiti 2010, Luhr 2011, Al-Sulaiti 2012}
Those studies primarily focused on fluence-correction factor, which is the ratio of dose in water to dose-to-water in material of interest at a same water-equivalent depth.
While the fluence-correction factor is essential for dosimetry with non-water phantom, generalization for a variety of modulated treatment beams may be difficult due to varied dose contribution from secondary particles and atomic recoils.

For carbon-ion radiotherapy, the radiation components need to be handled separately due to variability of relative biological effectiveness with particle charge and energy,\cite{Kase 2006} for which the fluence-correction factor for total dose would not suffice. 
Instead, survival of carbon ions that attenuate with depth to form a Bragg peak could be clinically more relevant.
In this regard, we evaluate water equivalence of some plastics for energy degrading of carbon-ion beams.

\section{Materials and methods}

\subsection{Sample plastics}

In addition to HDPE and PMMA, we theoretically evaluated two other oxygen-rich plastics, polyethylene terephthalate (PET) and polyoxymethylene (POM), among which we experimentally tested HDPE, PMMA, and POM with carbon-ion beams.
We excluded ABS as it varies in composition of carbon, hydrogen, and nitrogen. 
Table~\ref{tab:1} summarizes the relevant properties of these plastics and water,\cite{Groom 2001} in which the densities may vary with degree of polymerization.
Incidentally, $n_e = \langle Z/A_r \rangle \, \rho / u$ is the electron density, where $u$ is the atomic mass unit.

\begin{table}
\caption{\label{tab:1} Material properties (elemental composition, oxygen weight fraction $w_8$, typical density $\rho$, number of electrons per atomic mass unit $\langle Z/A_r \rangle$, and mean excitation energy $I$) excerpted from Ref.~\onlinecite{Groom 2001} (also available at \href{http://pdg.lbl.gov/}{the Particle Data Group website}) except for an empirical HDPE density.}
\setlength{\tabcolsep}{3pt}
\begin{tabular}{llcccc}
\hline\hline
Material & Composition & $w_8$ & $\rho / ({\rm g/cm^3}) $ & $\langle Z/A_r \rangle$ & $I/\textrm{eV}$ \\
\hline
HDPE & $({\rm C_2 H_4})_n$ & 0.000 & 0.96 & 0.57034 & 57.4 \\
PMMA & $({\rm C_5 H_8 O_2})_n$ & 0.320 & 1.19 & 0.53937 & 74.0 \\
PET & $({\rm C_{10} H_8 O_4})_n$ & 0.333 & 1.40 & 0.52037 & 78.7 \\
POM & $({\rm C H_2 O})_n$& 0.533 & 1.42 & 0.53287 & 77.4 \\
Water & ${\rm H_2 O}$ & 0.888 & 1.00 & 0.55509 & 79.7 \\
\hline\hline
\end{tabular}
\end{table}

\subsection{Effective densities for ion-beam Interactions}

\subsubsection{Stopping-power ratio}

Stopping power $S$ of a material for energetic charged particles is given by the Bethe theory.\cite{Cohen 2003} 
The effective density for energy degrading is the stopping-power ratio of material to water, which is defined as
\begin{eqnarray}
\frac{S}{S_w} = \frac{\langle Z/A_r \rangle \, \rho}{\langle Z/A_r \rangle_w \, \rho_w} \, \frac{\ln\frac{2 m_e c^2}{I} + \ln\frac{v^2}{c^2-v^2} -\frac{v^2}{c^2}}{\ln\frac{2 m_e c^2}{I_w} + \ln\frac{v^2}{c^2-v^2} -\frac{v^2}{c^2}},
\end{eqnarray}
where $\rho$, $\langle Z/A_r \rangle$, and $I$ for plastics and $\rho_w$, $\langle Z/A_r \rangle_w$, and $I_w$ for water are given in Table~\ref{tab:1}, $m_e$ is the electron mass, and $v$ and $c$ are the particle speed and the light speed in vacuum.
The $v$-dependence of $S/S_w$ is small for plastics with $I \approx I_w$.
For example, $S/S_w$ varies by $+0.9\%$ for slowing down in HDPE ($I/I_w = 72\%$) from $v^2/c^2 = 0.5$ ($E/A = 385.8$ MeV) to $v^2/c^2 = 0.1$ ($E/A = 50.4$ MeV), where $E/A$ is the kinetic energy per nucleon.
By representing the particle speed with $v^2/c^2 = 0.5$ in this study, the stopping-power ratio was approximated to be energy independent as
\begin{eqnarray}
\frac{S}{S_w} \approx \frac{\langle Z/A_r \rangle \, \rho}{\langle Z/A_r \rangle_w \, \rho_w} \, \frac{\ln\frac{2 m_e c^2}{I}-0.5}{\ln\frac{2 m_e c^2}{I_w}-0.5}, \label{eq:2}
\end{eqnarray}
on which the influence of $I$-value uncertainty is generally small.
For example, 3\% change of $I_w$ will cause only 0.3\% effect to $S/S_w$.
Incidentally, International Commission on Radiation Units and Measurements (ICRU) tentatively recommended $I_w = 78$ eV for water,\cite{Sigmund 2009} which deviates by $-1.7$ eV from the value in Table~\ref{tab:1}.

Alternatively, without depending on these uncertain $\rho$ and $I$ values, the stopping-power ratio can be directly measured by range shift $s$ in water per material thickness $t$ inserted upstream of water, 
\begin{eqnarray}
\frac{S}{S_w} \approx \frac{s}{t},
\end{eqnarray}
ignoring the air ($\rho/\rho_w \approx 0.1\%$) that is replaced by the inserted material.

\subsubsection{Scattering-power ratio}

The effective density for multiple scattering is the scattering-power ratio of material to water. 
Using the scattering-power formula for heavy charged particles by Gottschalk,\cite{Gottschalk 2010} it is defined as
\begin{eqnarray}
\frac{T}{T_w} = \frac{\rho}{91.69 \rho_w} \sum_i \frac{w_i}{{A_{\rm r}}_i} Z_i^2 \left( 29.73 - \ln Z_i - \ln {A_{\rm r}}_i \right),
\end{eqnarray}
where $Z_i$, ${A_r}_i$, and $w_i$ are the atomic number, the atomic weight, and the weight fraction of element $i$.

For range shift $s$, the increase of mean square angle of the primary particles in the material differs from that in water by a double-ratio factor,
\begin{eqnarray}
\Delta\overline{\theta^2}(s) = \frac{T/T_w}{S/S_w} \Delta\overline{\theta^2}_w(s). \label{eq:5}
\end{eqnarray}
The increase of mean square angle in water can be estimated by a semi-empirical formula,\cite{Kanematsu 2009}
\begin{eqnarray}
\Delta\overline{\theta^2}_w(s) = \frac{1.00}{1000}\, z^{-0.16} \left(\frac{m}{m_p}\right)^{-0.92} \ln \frac{R_0}{R_0-s}, \label{eq:6}
\end{eqnarray}
for particles of range $R_0$, charge $z e$, and mass $m$, where $e$ is the elementary charge and $m_p$ is the proton mass.

\subsubsection{Nuclear-cross-section ratio}

In the Sihver model,\cite{Sihver 1993} the geometrical cross section for collision between a projectile nucleus of mass number $A$ and a target nucleus of element $i$ is given by
\begin{eqnarray}
\sigma_{A i} = \pi r_0^2 \left[A^\frac{1}{3}+{A_r}_i^\frac{1}{3} 
-{b_0}_{A i} \left(A^{-\frac{1}{3}}+{A_r}_i^{-\frac{1}{3}}\right) \right]^2,
\end{eqnarray}
where $r_0 = 1.36$ fm is the effective nucleon radius and $b_0$ is the transparency parameter, for which the proton--nucleus formula is applied for collisions on hydrogen ($i=1$), namely
\begin{eqnarray}
{b_0}_{A i} = \begin{cases}
2.247- 0.915 \left(A^{-\frac{1}{3}}+{A_r}_i^{-\frac{1}{3}}\right) & \text{for } i =1 \\
1.581- 0.876 \left(A^{-\frac{1}{3}}+{A_r}_i^{-\frac{1}{3}}\right) & \text{for } i > 1.
\end{cases}
\end{eqnarray}
The effective density for nuclear interactions is the nuclear-cross-section ratio of material to water, \begin{eqnarray}
\frac{\sigma_A}{{\sigma}_{A w}} =\frac{ \langle \sigma_A/A_r \rangle \, \rho} {\langle \sigma_A/A_r \rangle_w \,\rho_w}, 
\end{eqnarray}
where 
\begin{eqnarray}
\langle \sigma_A/A_r \rangle = \sum_i \frac{w_i}{{A_r}_i} \sigma_{A i}
\end{eqnarray}
is the nuclear cross section per atomic mass unit of the material.
The energy dependence of the nuclear-cross-section ratio may be reasonably ignored due to cancellation of common energy dependence for $E/A \gtrsim 100$ MeV expected in the Sihver model.

Number of primary particles $N$ decreases in matter due to nuclear interactions.
With an insert for range shift $s$, the number at depth $d$ in water is factorized as
\begin{eqnarray}
N(d) = \alpha(s)\, N_w(s+d),
\end{eqnarray}
where $N_w(d)$ is the number of primary particles at depth $d$ in water for $s = 0$ and survival ratio $\alpha(s) = N(0)/N_w(s)$ is the ratio of the number of carbon ions after range shift $s$ to that in water at depth $d = s$.
The fractional attenuation per range shift, $(-dN/ds)/N$, differs between the material and water by a double-ratio factor,
\begin{eqnarray}
-\frac{1}{N}\frac{dN}{ds} = \frac{\sigma_A/\sigma_{A w}}{S/S_w} \left(-\frac{1}{N}\frac{dN}{ds}\right)_w,
\end{eqnarray}
which is solved with boundary condition $N= N_w$ for $s = 0$, resulting in 
\begin{eqnarray}
\alpha(s) = \left(\frac{N_w(d=s)}{N_w(d=0)}\right)^{\frac{\sigma_A/\sigma_{A w}}{S/S_w}-1}.
\end{eqnarray}

If we employ an empirical formula for attenuation of carbon ions in water described in Appendix with substitution of residual range $R = R_0-s$, the survival ratio in the exponential region ($s \leq R_0-2$~cm) reduces to
\begin{eqnarray}
\alpha(s) = \exp \left( -\frac{s}{25.5~\rm cm}\left(\frac{\sigma_{12}/\sigma_{12 w}}{S/S_w}-1\right) \right), 
\label{eq:14}
\end{eqnarray}
which analytically gives the carbon-ion survival in the plastic relative to that in water.

\subsection{Experiment with carbon-ion beams}

The dosimetric water equivalence of different materials should be evaluated by equality of the resultant doses.
As far as the primary particles are concerned, the Bethe theory for energy loss and the Moli\`ere theory for multiple scattering are accurate to the level of measurement limit if only the $\rho$ and $I$ values are adjusted to a specific material.\cite{Cohen 2003, Gottschalk 2010}
On the contrary, the nuclear-interaction models generally use assumptions, approximations, and extensive cross-section data with finite uncertainties and limitations.
Although the model used in this study has been tested,\cite{Sihver 2009} specific validation may be necessary for this particular application.

We conducted an experiment with carbon-ion beams of $E/A = 290$ and 430 MeV extracted from Heavy Ion Medical Accelerator in Chiba (HIMAC) of National Institute of Radiological Sciences.
The beams were laterally broadened by a wobbling/scattering system to form a uniform field of 10-cm diameter,\cite{Torikoshi 2007} longitudinally moderated by a ripple filter for Gaussian range modulation of 1.8 mm (rms),\cite{Schaffner 2000} and delivered horizontally to a box-shaped water tank on a movable treatment couch.

\begin{figure}
\includegraphics{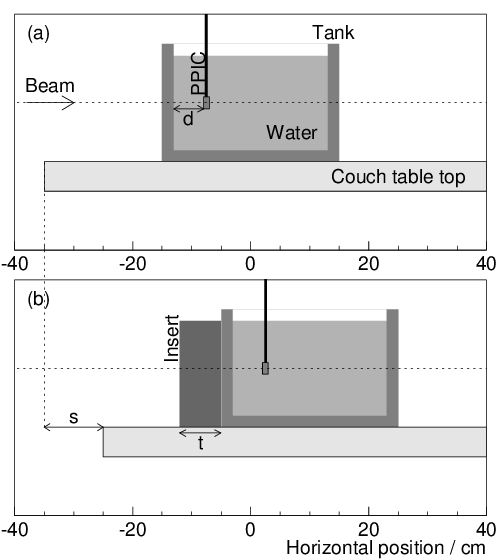}
\caption{\label{fig:1} Side views of apparatus for carbon-ion-beam experiment, (a) for the reference condition and (b) with a plastic insert of thickness $t$ and couch movement by range shift $s$, where the central-axis (horizontal-dotted-line) doses at varied depth $d$ were measured with PPIC.}
\end{figure}

We used binary plates of HDPE, PMMA, and POM, which were stacked at the immediate upstream of the tank to construct an insert of arbitrary thickness $t$ in steps of 1 cm.
To eliminate the influence of beam divergence for wobbling at about 10 m upstream, the treatment couch was moved downstream so that the Bragg peak would stay in the same place in the laboratory system.
The amount of couch movement was the expected range shift $s = (S/S_w)\,t$ using our standard $S/S_w$ values 1.02 for HDPE, 1.16 for PMMA, and 1.36 for POM.
The central-axis doses at varied depth $d$ in water after the tank wall of 19-mm PMMA were measured with a Markus-type (PTW Type 23343) plane-parallel ionization chamber (PPIC) with a protective cover of 0.87-mm PMMA, as shown in Fig.~\ref{fig:1}.
The dosimetric precisions were as good as 0.1\% for peak dose and 0.1 mm for depth in reproducibility.

We chose insert thicknesses that approximately corresponded to range shifts of 4 cm and 8 cm for 290 MeV and 10 cm and 20 cm for 430 MeV.
For beam deliveries of equal monitor units, the central-axis doses were measured at minimum depth intervals of 0.1 mm.
For the reference depth of the moderated Bragg peak, we adopted depth $d_{80}$, at which the dose decreased to 80\% of peak dose $D_p$.
Range shift $s$ was measured as the shift of $d_{80}$ by insertion.
Using the dose points in the depth region of $d_{80}+1 ~{\rm cm} \lesssim d \lesssim d_{80}+2~ {\rm cm}$, we estimated the fragment dose $D_f$ by extrapolation of the linear-fit line to $d = d_{80}$, for which we assumed the linearity over a few centimeters and ignored small dose variation over about 2 mm between the peak and 80\% depths.
Taking the beam without insert ($s = 0$) as a reference, the carbon-ion-survival ratios were estimated by reduction of the carbon-ion dose contribution,
\begin{eqnarray}
\alpha(s) = \frac{D_p(s)-D_f(s)}{D_p(0)-D_f(0)},
\end{eqnarray}
which were then compared with the analytical model calculations.

\section{Results}

\subsection{Theoretical effective densities}

Table~\ref{tab:2} shows the resultant effective-density calculations in a format insensitive to $\rho$ variation.
The dosimetric water equivalence, that is the uniformity of effective densities for the relevant interactions, was highly correlated with the oxygen content of these plastics.
For HDPE of $\rho = 0.96$ g/cm$^3$, although its electron-density ratio $n_e/ {n_e}_w = 0.985$ and stopping-power ratio $S/S_w = 1.020$ are close to 1, it differs from water by $-29.0\%$ in multiple scattering and $+11.0\%$ in nuclear interactions.
Equation~(\ref{eq:14}) leads to the fractional attenuation ratios per range shift of $(-d\alpha/ds)/\alpha =$ 0.45\%/cm for HDPE, 0.17\%/cm for PMMA, 0.04\%/cm for PET, and 0.02\%/cm for POM, which indicate the differences from water in attenuation.

\begin{table}
\caption{\label{tab:2} Theoretical double ratios of plastics to water (with subscript $w$) between density $\rho$, electron density $n_e$, stopping power $S$, scattering power $T$, and carbon-ion nuclear cross section $\sigma_{12}$.}
\setlength{\tabcolsep}{3pt}
\begin{tabular}{lcccc}
\hline\hline
Material & $\frac{n_e}{{n_e}_w}/\frac{\rho}{\rho_w}$ & $\frac{S}{S_w}/\frac{\rho}{\rho_w}$ & $\frac{T}{T_w}/\frac{S}{S_w}$ & $\frac{\sigma_{12}}{\sigma_{12 w}}/\frac{S}{S_w}$\\
\hline
HDPE & 1.027 & 1.065 & 0.712 & 1.114 \\
PMMA & 0.972 & 0.980 & 0.890 & 1.044 \\
PET & 0.937 & 0.939 & 0.953 & 1.011 \\
POM & 0.960 & 0.963 & 0.968 & 1.004 \\
\hline\hline
\end{tabular}
\end{table}

\subsection{Experimental dose variation}

\begin{figure}
\includegraphics{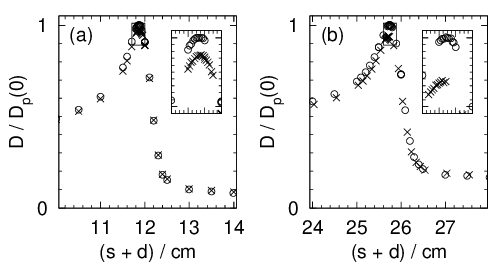}
\caption{\label{fig:2} Depth--dose distributions of the moderated carbon-ion Bragg peak with ($\times$) and without ($\bigcirc$) insertion of (a) 8-cm HDPE for 290 MeV and (b) 20-cm HDPE for 430 MeV, with embedded magnified plots.}
\end{figure}

Figure~\ref{fig:2} shows explicit cases for variation of the moderated Bragg peak by material insertion, with which the peak and fragment doses were measured.
The resultant measurements are shown in Table~\ref{tab:3}, where available range $R_0$ is a sum of the effective thickness of the tank wall, that of the PPIC cover, and the range-equivalent depth measured for $s = 0$.
For proton beams with large range straggling, 80\%-dose depth $d_{80}$ is commonly used as the range-equivalent depth.\cite{Bortfeld 1997} 
For the moderated Bragg peak formed by the stopping carbon ions, fragment contribution $D_f/D_p$ was excluded to redefine the range-equivalent depth with modified relative dose $(1-D_f/D_p)\times 80\% + D_f/D_p$, which was 82\% for 290 MeV and 84\% for 430 MeV.

\begin{figure}
\includegraphics{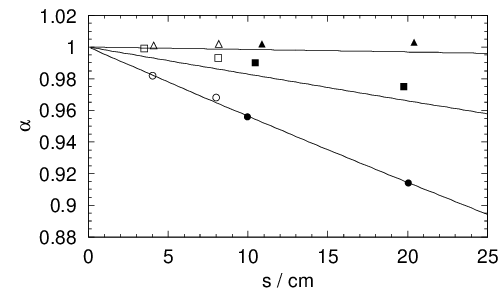}
\caption{\label{fig:3} Carbon-ion survival ratio $\alpha$ measured as Bragg-peak reduction for range shift $s$ by HDPE (circles), PMMA (squares), and POM (triangles) for 290-MeV (open symbols) and 430-MeV (filled symbols) beams along with analytical model curves (solid lines).}
\end{figure}

Figure~\ref{fig:3} shows the measured and calculated survival ratios for range shift by these plastics.
Invariance of the Bragg peak is essential for treatment planning and delivery systems based on water-phantom dosimetry.
The reduction of the fragment-subtracted Bragg peak was 8.9\% with HDPE for 20-cm range shift to a carbon-ion beam of 28.23-cm range while it was 2.5\% with PMMA and 0.0\% with POM.
The dominant source of uncertainty may be the fragment dose estimated by linear extrapolation of tail doses sampled at a few depths, which would have caused $ \lesssim 1\%$ effect to the subtracted peak dose.
In fact, the analytical model calculations reasonably agreed with the measurements within 1\% in survival ratio.

\begin{table} 
\caption{\label{tab:3} Experimental results for carbon-ion attenuation by plastic inserts, showing acceleration energy $E/A$, available range $R_0$, material thickness $t$, range shift $s$, 80\%-dose depth $d_{80}$, relative peak and fragment doses $D_p$ and $D_f$, and carbon-ion-survival ratio $\alpha$.}
\setlength{\tabcolsep}{3pt}
\begin{tabular}{lcccccc}
\hline\hline
\multicolumn{3}{l}{$E/A = 290$ MeV} & \multicolumn{3}{l}{$R_0 = 14.36$ cm} \\
Material & $t$/cm & $s$/cm & $d_{80}$/cm & $D_p$ & $D_f$ & $\alpha$ \\
\hline
(Reference) & 0 & 0 & 12.06 & 1 & 0.115 & 1 \\
\hline
HDPE & 4 & 4.02 & ~8.04 & 0.987 & 0.118 & 0.982 \\
& 8 & 8.00 & ~4.06 & 0.974 & 0.117 & 0.968 \\
\hline
PMMA & 3 & 3.48 & ~8.58 & 0.998 & 0.114 & 0.999 \\
& 7 & 8.14 & ~3.92 & 0.992 & 0.113 & 0.993 \\
\hline
POM & 3 & 4.09 & ~7.97 & 1.000 & 0.114 & 1.001 \\
& 6 & 8.17 & ~3.89 & 1.001 & 0.114 & 1.002 \\
\hline\hline
\multicolumn{3}{l}{$E/A = 430$ MeV} & \multicolumn{3}{l}{$R_0 = 28.23$ cm} \\
Material & $t$/cm & $s$/cm & $d_{80}$/cm & $D_p$ & $D_f$ & $\alpha$ \\
\hline
(Reference) & 0 & 0 & 25.95 & 1 & 0.199 & 1 \\
\hline
HDPE & 10 & ~9.97 &15.98 & 0.966 & 0.200 & 0.956 \\
& 20 & 20.04 & ~5.91& 0.937 & 0.205 & 0.914 \\
\hline
PMMA & ~9 & 10.45 & 15.50 & 0.992 & 0.199 & 0.990 \\
& 17 & 19.75 & ~6.20 & 0.981 & 0.200 & 0.975 \\
\hline
POM & ~8 & 10.88 & 15.07 & 1.001 & 0.198 & 1.002 \\
& 15 & 20.42 & ~5.53 & 0.999 & 0.195 & 1.003 \\
\hline\hline
\end{tabular}
\end{table}

\section{Discussion}

As the density of polymer plastics depends on manufacturing condition, the effective density must be experimentally determined before its clinical use to an accuracy better than 1\%.
Therefore, the stopping-power similarity to water ($S \approx S_w$) for HDPE is not essentially an advantage.
The similarity is only relevant when geometrical dose properties are of interest, such as field width and penumbra size.
For applications that allow or require regular air gaps in a phantom, higher $S/S_w$ values are acceptable or may even be preferable.
For example, Yajima {\it et al.} developed a multilayer ionization chamber composed of 3-mm and 4-mm PMMA plates interleaved with 1-mm air gaps,\cite{Yajima 2009} 
which would reduce the density by factor $7/9$ resulting in mean stopping-power ratio $\langle S/S_w\rangle = 0.90$.
If POM is used instead, the $\langle S/S_w\rangle$ value will be 1.06, which will further be adjusted to 1.00 by extending the air gaps to 1.26 mm.

Beam blurring due to multiple scattering in a range compensator should ideally be minimized.
In this regard, HDPE with $(T/T_w)/(S/S_w) = 0.712$ is superior to PMMA (0.890), PET (0.953), POM (0.968), and water (1) although their effective differences will be generally marginal.
For example, when a carbon-ion beam of 20-cm available range traverses a plastic plate for 10-cm range shift, the induced beam blurring in 30-cm distance will be 1.9 mm for HDPE, 2.1 mm for PMMA, 2.2 mm for PET, and 2.2 mm for POM, using Eqs.~(\ref{eq:5}) and (\ref{eq:6}). 

The dosimetric water equivalence of the plastics that we tested here originated from the similarity with water in oxygen content and should be also valid for protons and other ions.
Therefore, POM will be generally a good dosimetric material if its high density is tolerable.
For some dosimetric applications that require material transparency, such as water tank with optical dosimeter-alignment system, PET may desirably substitute for PMMA.

\section{Conclusions}

We evaluated dosimetric water equivalence of four common plastics, HDPE, PMMA, PET, and POM, by uniformity of effective densities for carbon-ion-beam interactions.
Among them, POM was the best and virtually indistinguishable from high-density water, which would be ideal for range control and preferable for dosimetry with regular air gaps such as for multilayer ionization chamber.
For applications that require transparency such as water tank with optical dosimeter-alignment system, PMMA was verified to be reasonably water equivalent and PET would be even better.
The poorest was HDPE with a large fractional attenuation ratio of 0.45\% per 1-cm range shift, although with marginal superiority for reduced multiple scattering.
Analytical model calculations agreed with measurements within 1\% for carbon-ion-survival ratio of plastic to water.
The water equivalence of plastics was highly correlated with their oxygen content, as expected from the composition of water.

\appendix*\section{Attenuation of carbon ions in water}

Haettner {\it et al.} measured the number of carbon ions, $N_w$, attenuating with depth $d$ in water for $E/A = 200$ MeV and 400 MeV.\cite{Haettner 2006}
We compiled their data to construct an energy-independent universal formula as a function of residual range in water, $R = d_R-d$, where $d_R$ is the range-equivalent depth.

Using the measured $N_w(d)$ curve for each energy, the $d_R$ was tentatively set to the maximum-gradient depth.
Using the data points for $d_R+0.5~{\rm cm} < d < d_R+2~{\rm cm}$, a straight line was fitted and its intercept at $d=d_R$ was defined as the number of stopped carbon ions, $N_{w R}$.
The $d_R$ was then redefined with submillimeter adjustment to meet $N_w(d_R) = N_{w R}/2$.

\begin{figure}
\includegraphics{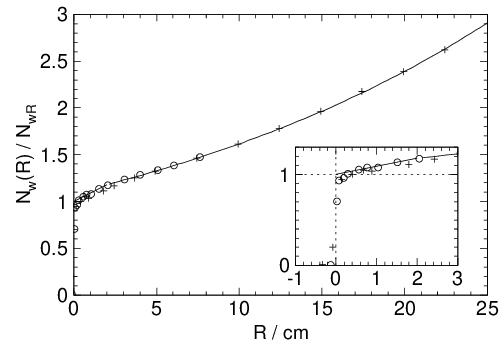}
\caption{\label{fig:4} Number of carbon ions, $N_w$, per number of stopped carbon ions, $N_{w R}$, as a function of residual range in water, $R$, measured by Haettner {\it et al.}\cite{Haettner 2006} for $E/A = 200$ MeV ($\bigcirc$) and 400 MeV ($+$) with the fitted function (solid line).}
\end{figure}

The measurements showed natural exponential behavior for $R > 5$ cm, for which we determined the mean free path using the 400-MeV dataset.
While there would naturally be some difference between the energies in the small region of $R < 1$ cm due to marginal variation of range straggling, systematic deviation was observed unexpectedly in the larger region of $R < 5$ cm.
As the exponential behavior held in the intermediate region of $2~{\rm cm} < R < 5~{\rm cm}$ for the 200-MeV dataset, the deviation may be attributed to increased measurement difficulty for the 400-MeV beam.
Therefore, we rescaled the 400-MeV dataset to match the 200-MeV dataset in the exponential region of $R > 5$ cm and we adopted the straight line fitted to the 200-MeV dataset for the $R \leq 2$ cm region.
The resultant universal formula for carbon-ion attenuation is
\begin{eqnarray}
\frac{N_w(R)}{N_{w R}} = 
\begin{cases}
1+\cfrac{R}{11.1~{\rm cm}} & \text{for } R \leq 2~{\rm cm} \\
1.091 \exp\left(\cfrac{R}{25.5~{\rm cm}}\right) & \text{for } R > 2~{\rm cm},
\end{cases}
\end{eqnarray}
which is shown in Fig.~\ref{fig:4} along with the measurements rescaled as described above.

\end{document}